\documentclass[prd, amsfonts, onecolumn, nofootinbib, showpacs]{revtex4-2}
\usepackage{graphicx}
\usepackage{epsfig}
\usepackage{color}
\usepackage{amsmath}
\usepackage{amssymb}
\usepackage{lineno}
\newcommand{\be}{\begin{equation}}
\newcommand{\ee}{\end{equation}}
\newcommand{\bea}{\begin{eqnarray}}
\newcommand{\eea}{\end{eqnarray}}

\newcommand{\gapp}{\mathrel{\raise.3ex\hbox{$>$}\mkern-14mu
\lower0.6ex\hbox{$\sim$}}}
\newcommand{\lapp}{\mathrel{\raise.3ex\hbox{$<$}\mkern-14mu
\lower0.6ex\hbox{$\sim$}}}
\def\bbox{{\,\lower0.9pt\vbox{\hrule \hbox{\vrule height 0.2 cm
\hskip 0.2 cm \vrule  height 0.2 cm}\hrule}\,}}

\begin{document}
\title{On black holes as macroscopic quantum objects}
%\author{}
%\affiliation{ }
\author{De-Chang Dai$^{1,4}$\footnote{communicating author: De-Chang Dai,\\ email: diedachung@gmail.com\label{fnlabel}}, Djordje Minic$^2$, Dejan Stojkovic$^3$}
\affiliation{$^1$ Center for Gravity and Cosmology, School of Physics Science and Technology, Yangzhou University, 180 Siwangting Road, Yangzhou City, Jiangsu Province, P.R. China 225002 }
\affiliation{ $^2$ Department of Physics, Virginia Tech, Blacksburg, VA 24061, U.S.A. }
%\affiliation{ $^3$ }
\affiliation{ $^3$ HEPCOS, Department of Physics, SUNY at Buffalo, Buffalo, NY 14260-1500, U.S.A.}
\affiliation{ $^4$ CERCA/Department of Physics/ISO, Case Western Reserve University, Cleveland OH 44106-7079}

 %%%%%%%%%%%%%%%%%%%%%%%%%%%%%%%%%%%%%%%%%%%%%%%%%%%%%%%

\begin{abstract}
\widetext
The relative flow of the Schwarzschild vs. the proper time during the classical evolution of a collapsing shell in the Schwarzschild coordinates practically forces us to interpret black hole formation as a highly non-local quantum process in which a shell/anti-shell pair is created within the incipient horizon, thus canceling out the original collapsing shell exactly at the horizon. By studying quantum fields in the black hole background, we reveal similar non-local effects. Among other things, the outgoing member of the Hawking pair very quickly becomes entangled with the black hole geometry (and not its partner), which is in contrast with the usual assumption that the  Hawking pair is maximally entangled according to the local geometry near the horizon. Also, an infalling wave affects the black hole geometry even before it crosses the horizon.
Finally, we find that a particle takes a finite amount of time to tunnel in and out of the black hole horizon, and thus avoids infinite blue and redshift in processes happening exactly at the horizon. These findings strongly support the picture of a black hole as a macroscopic quantum object.   
\end{abstract}

%%%%%%%%%%%%%%%%%%%%%%%%%%%%%%%%%%%%%%%%%%%%%%%%%%

\maketitle

\section{ Introduction and overview}

Black holes are among the most fascinating objects in physics and astronomy \cite{frolov}.
They are also believed to hold some of the most important secrets of quantum gravity, perhaps
the most outstanding problem in theoretical physics.
In this letter we offer a new perspective of the nature of black holes.
Our main point is that even though black holes are undoubtedly classical solutions of general relativity,
they can be also understood as macroscopic quantum objects.
We present explicit and concrete calcuations that support this, perhaps surprising, point of view.

\section{ ``Classical'' black hole formation viewed as a macroscopic quantum process }
\label{BHF}

We consider the gravitational collapse of a massive shell of radius $R(t)$. According to Birkhoff's theorem, the metric inside the shell, for $r<R(t)$, is the Minkowski metric
\begin{equation}
ds^2 =  -dT^2 +dr^2 +r^2 d\Omega^2,
\end{equation}
while outside, for $r>R(t)$, is the Schwarzschild metric
\begin{equation}
\label{SW1}
ds^2 = - \left(1-\frac{2M}{r}\right)dt^2 +\left(1-\frac{2M}{r}\right)^{-1}dr^2 +r^2 d\Omega^2,
\end{equation}
where for simplicity we set $G=1$.

\begin{figure}
   %\centering
\includegraphics[width=8cm]{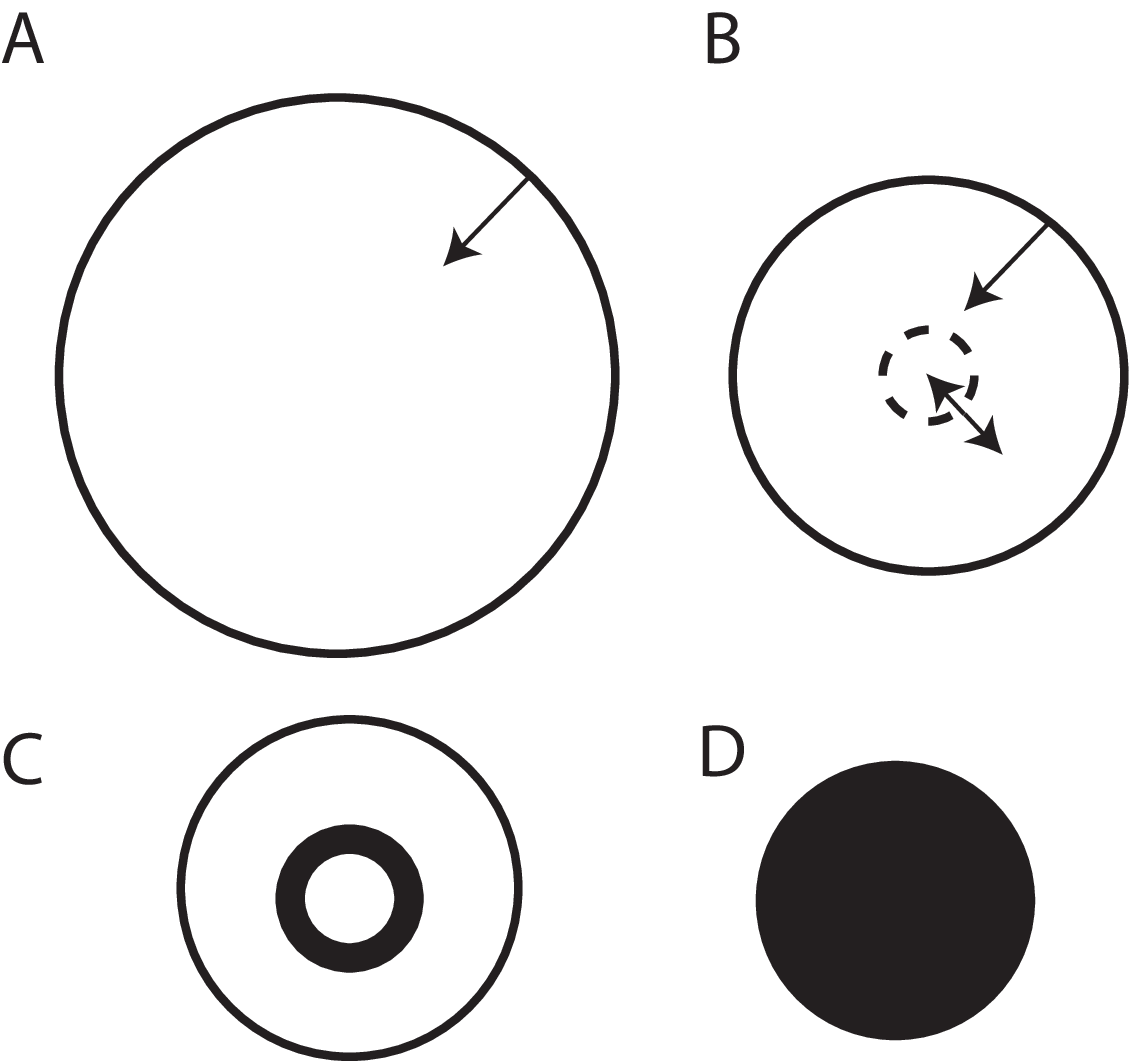}
\caption{ {\bf A.} The schematic representation of a collapsing spherical shell. The metric is  
Schwarzschild outside, and Minkowski  inside. {\bf B.} The shell/anti-shell pair is created with the radius  $R=\frac{\mu^2}{2M}$. The positive energy shell falls to $r=0$, while the negative energy one propagates outward to the incipient horizon. {\bf C.} The region between the created shells (black ring) is growing, and it is not a flat space anymore. Its time coordinate is re-synchronized into the Schwarzschild time. {\bf D.} Eventually, the negative energy shell cancels out the original collapsing shell. A black hole is formed and the whole spacetime becomes Schwarzschild. } 
\label{mass}
\end{figure}

The explicit motion of the shell can be found (see e.g. Lightman et al. \cite{grbook}, Problems 21.10 and 21.11) from the conserved quantity, $M$, which is just the total energy of the shell 
\begin{equation}
\label{motion1}
M=\mu \sqrt{1+\dot{R}^2} -\frac{\mu^2}{2R}
\end{equation}
where $\dot{R}=dR/d\tau $, $\tau$ is the proper time of an observer sitting on the shell, and $\mu$ is the rest mass of the shell. While the evolution in terms of the proper time $\tau$ is uneventful  (the shell shrinks to $R=0$ in finite time), it is instructive to see what happens in the Schwarzschild time. The relative flow of the proper and Schwarzschild times during the motion of the shell can be found from the time component of the four-velocity $u^t$ (see detailed derivation in the appendix)
\begin{equation} \label{ut}
 u^t = \frac{dt}{d\tau}=
  \begin{cases} 
   \frac{(1-\frac{2M}{R}+\dot{R}^2)^{1/2}}{1-\frac{2M}{R}} ,  & \text{if } R > \frac{\mu^2}{2M} \\
   -\frac{(1-\frac{2M}{R}+\dot{R}^2)^{1/2}}{1-\frac{2M}{R}} ,  & \text{if } R < \frac{\mu^2}{2M} .
  \end{cases}
\end{equation}
This relation is crucial for our discussion. 
For $R>2M$, $u^t>0$, which means that the Schwarzschild and proper infalling time coordinates are of the same sign. The shell propagating according to the Schwarzschild coordinate time behaves as a normal positive energy particle. However, for $2M>R>\frac{\mu^2}{2M}$, $u^t<0$. Thus, time is reversed and the shell behaves as a negative energy particle. Time reverses once again for $\frac{\mu^2}{2M}>R$, where $u^t>0$. The shell again behaves as a positive energy  particle. This behavior can be interpreted as a shell/anti-shell pair creation with the radius $R=\frac{\mu^2}{2M}$. The positive energy member of the pair travels to $r=0$, and presumably forms a singularity there. The negative energy member travels to the horizon and cancels the incoming positive mass shell. The region between the created shells is not flat anymore, and time is re-synchronized into the Schwarzschild time. Eventually, the whole spacetime becomes Schwarzschild-like, and an outside observer does not see the infalling shell anymore.   The schematics is shown in Fig.~\ref{mass}. 

To corroborate this description, we calculate the detailed trajectory of the shell in the Schwarzschild coordinates by integrating 
\begin{equation} \label{ts}
t=\int \frac{u^t}{\dot{R}} dR  ,
\end{equation} 
where $\dot{R}$ is given by Eq.~(\ref{motion1}), and $u^t$  by Eq.~(\ref{ut}).
In order to make the relevant plots we set $M=\mu=\frac{1}{2}$. Since both $M$ and $\mu$ are conserved quantities, this choice corresponds to the shell which starts from rest at infinity.  The explicit trajectory in Schwarzschild coordinates is 
\begin{equation} \label{tsch}
 t=
  \begin{cases} 
  - \frac{R+2}{3}\sqrt{8R+1}-\ln(\frac{\sqrt{8R+1}-3}{\sqrt{8R+1}+3})& \text{if } R > 1 \\
   - \frac{R+2}{3}\sqrt{8R+1}-\ln(\frac{3-\sqrt{8R+1}}{\sqrt{8R+1}+3}) & \text{if } 1>R  .
  \end{cases}
\end{equation}

\begin{figure}
   %\centering
\includegraphics[width=10cm]{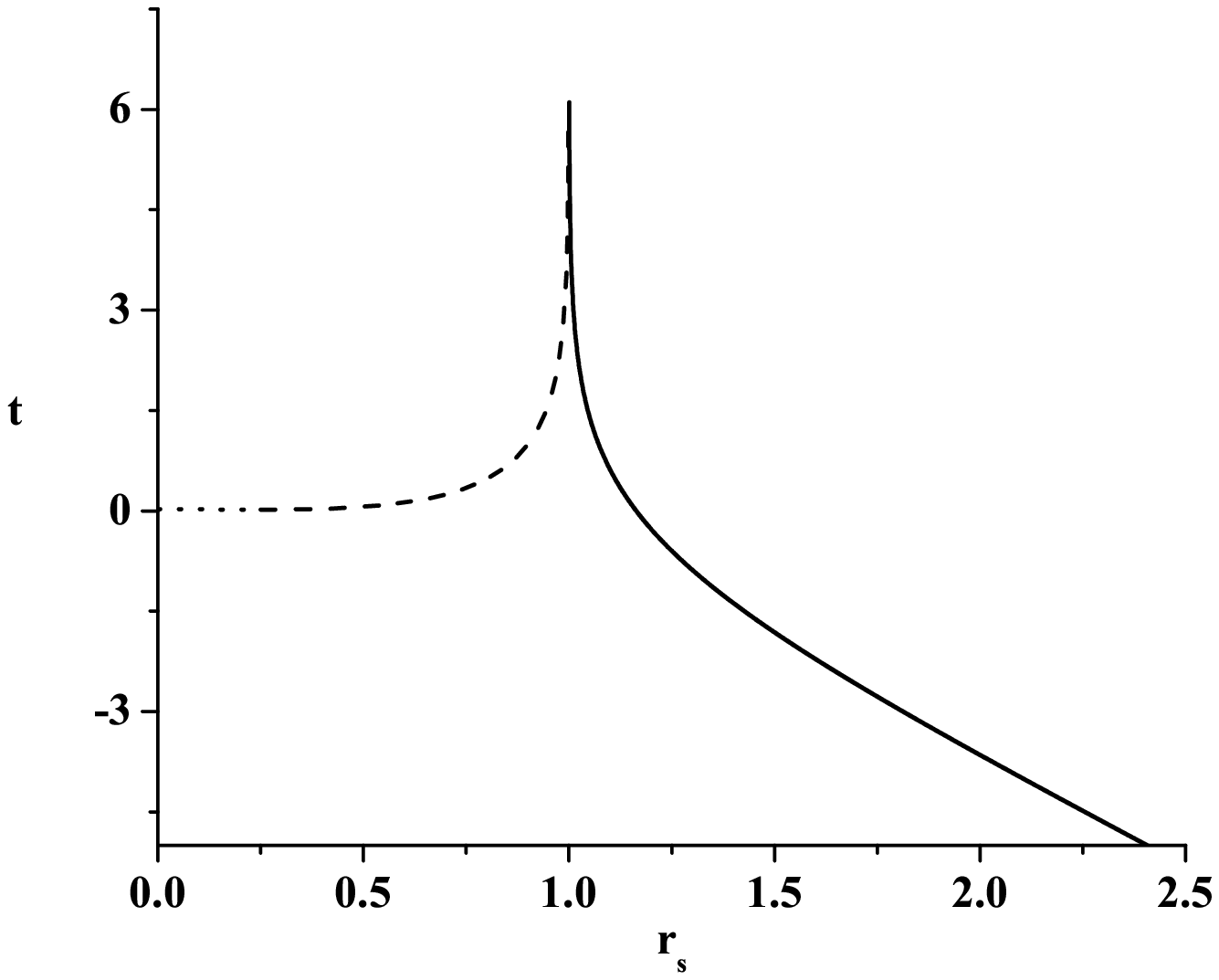}
\caption{The trajectory of a collapsing shell in the Schwarzschild coordinates $(t,r)$. We set the rest mass of the shell, $\mu$, and its total energy, $M$, to be equal to $1/2$. The incipient black hole horizon is at $r=1$. The collapsing shell takes infinite time to arrive to the horizon (solid line). However, a shell/anti-shell pair is created with the radius $R=0.25$. The positive energy shell shrinks to $R=0$ (dotted line), while the negative energy one propagates outward (the dashed line), where finally it cancels the original collapsing shell at the horizon at $t\rightarrow \infty$.   } 
\label{massshell}
\end{figure}

 As Fig.~\ref{massshell} shows, there is only one mass shell in the beginning. When the shell reaches $R\approx 1.16 r_h$, where $r_h$ is the horizon radius, the shell/anti-shell pair is created with the radius $R=0.5\mu$. The positive energy shell keeps falling into $r=0$, while the negative energy shell proceeds toward the incipient horizon.  Eventually, the outgoing negative energy shell reaches the horizon where it cancels out the original infalling shell. 
The regions above these curves are Schwarzschild-like, while below, they are flat. 

This result is remarkable for two reasons. First, we work in the framework of classical general relativity. Yet, we are practically forced to interpret the process of the collapse in quantum mechanical terms as a shell/anti-shell creation. Second, while the original collapsing shell is still outside its own Schwarzschild radius, the region inside is already affected in a highly non-local way. We emphasize that this is the description of the collapse in the Schwarzschild coordinates. An infalling observer on the shell who measures the proper time $\tau$ will hit the singularity in finite time according to his clock.

\section{ Quantum fields in the background of a black hole~ }\label{QF}

\begin{figure}
   %\centering
\includegraphics[width=10cm]{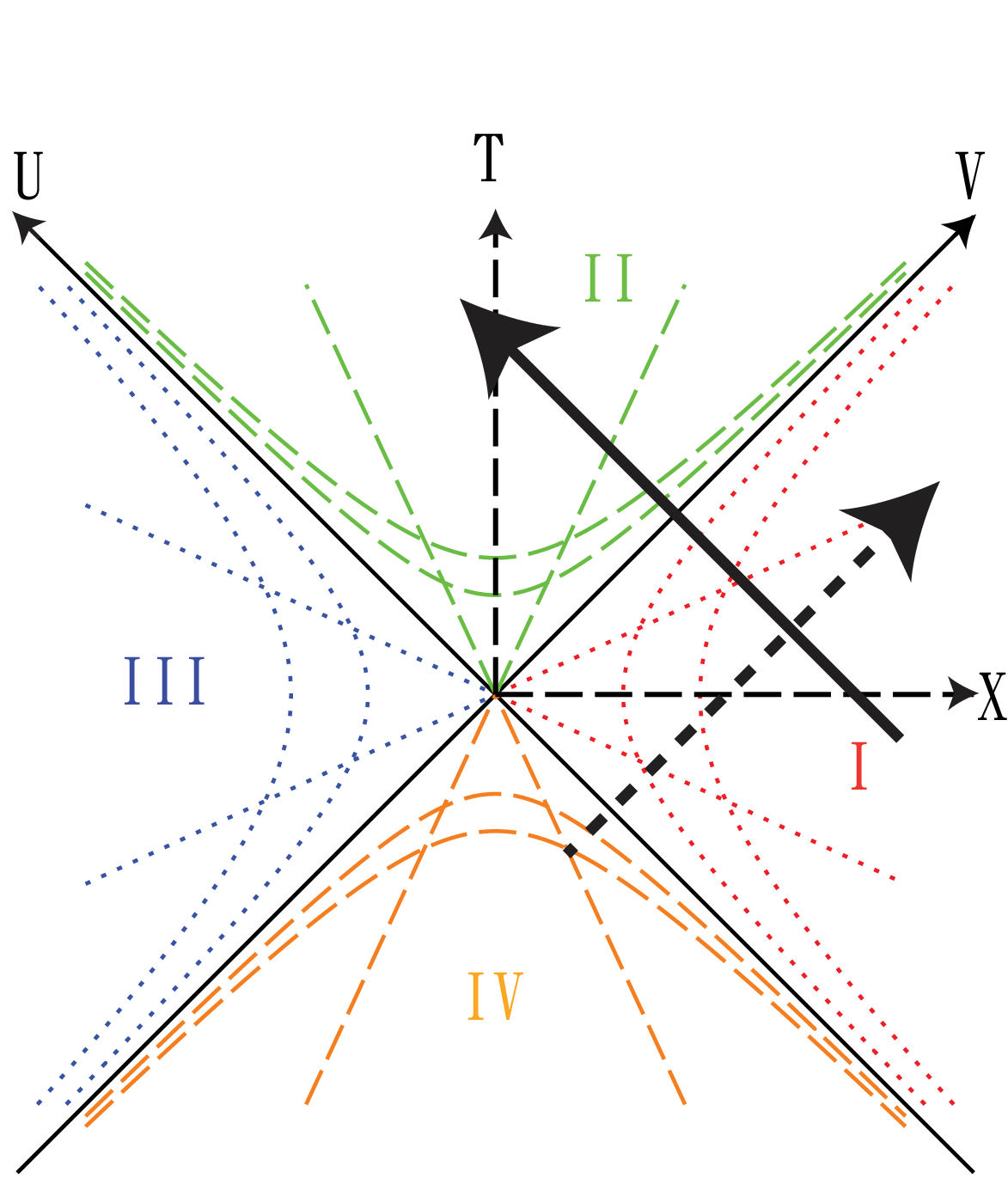}
\caption{A Schwarzschild black hole in the Kruskal-Szekeres coordinates: Region I is outside, while region II is inside the horizon. Regions III and IV are copies of regions I and II. The $U$ and $V$ axes are along the horizon. The dashed and solid arrow lines represent constant $U$ and $V$ trajectories.  The other lines represent constant $r$ and $t$ trajectories.    
}
\label{kruskal}
\end{figure}

We will now introduce a quantum field in the background of a black hole, and study what happens to the infalling and outgoing waves.  In this case, the black hole is already formed, and the metric is given by Eq.~ (\ref{SW1}). To make the relevant plots we set $2M=1$. To remove the coordinate singularity at the horizon, $r=1$, we introduce the Kruskal-Szekeres coordinates $(T,X)$. For $r>1$, we have  
\begin{eqnarray}
\label{outerT}
T&=&\Big(r-1\Big)^{1/2} e^{r/2} \sinh\Big(t/2\Big)\\
\label{outerX}
X&=&\Big(r-1\Big)^{1/2} e^{r/2} \cosh\Big(t/2\Big),
\end{eqnarray}
 while for $r<1$,  
\begin{eqnarray}
\label{innerT}
T&=&\Big(1-r\Big)^{1/2} e^{r/2} \cosh\Big(t/2\Big)\\
\label{innerX}
X&=&\Big(1-r\Big)^{1/2} e^{r/2} \sinh\Big(t/2\Big).
\end{eqnarray}
 Note that Eqs. \eqref{outerT} and \eqref{outerX} are written for the quarter I in Fig. \ref{kruskal}, while Eqs. \eqref{innerT} and \eqref{innerX} are written for the quarter II in Fig. \ref{kruskal}. There is an extra negative sign in these expression for the quarters III and IV.
We can also replace $(T, X)$ with the lightcone coordinates $U$ and $V$ as  
$U=T-X$ and $V=T+X$.
%\begin{eqnarray}
%U&=&T-X\\
%V&=&T+X.
%\end{eqnarray}
The metric in Eq.~(\ref{SW1}) is now written as 
\begin{equation}
ds^2 =  \frac{4}{r}e^{-r} dUdV +r^2 d\Omega^2.
\end{equation}
For simplicity, we can omit the angular part of the metric and consider the $1+1$ dimensional space 
\begin{equation}
ds^2 =  \frac{4}{r}e^{-r} dUdV.
\end{equation}

A massless scalar field propagating in this background must satisfy the $1+1$-dimensional Klein-Gordon equation
\begin{equation}
\partial_U\partial_V \Phi=0.
\end{equation}  
The solution of this equation can be written as a linear combination of two functions, $f(U)$ and $g(V)$, 
%\begin{equation}
$\Phi =A f(U) +B g(V),$
%\end{equation}
where A and B are constants. The solution can be transformed to Schwarzschild coordinates by substituting the explicit forms for $T(t,r)$ and $X(t,r)$ obtained from Eqs. (\ref{outerT}) - (\ref{innerX}).

\begin{figure}
   %\centering
\includegraphics[width=10cm]{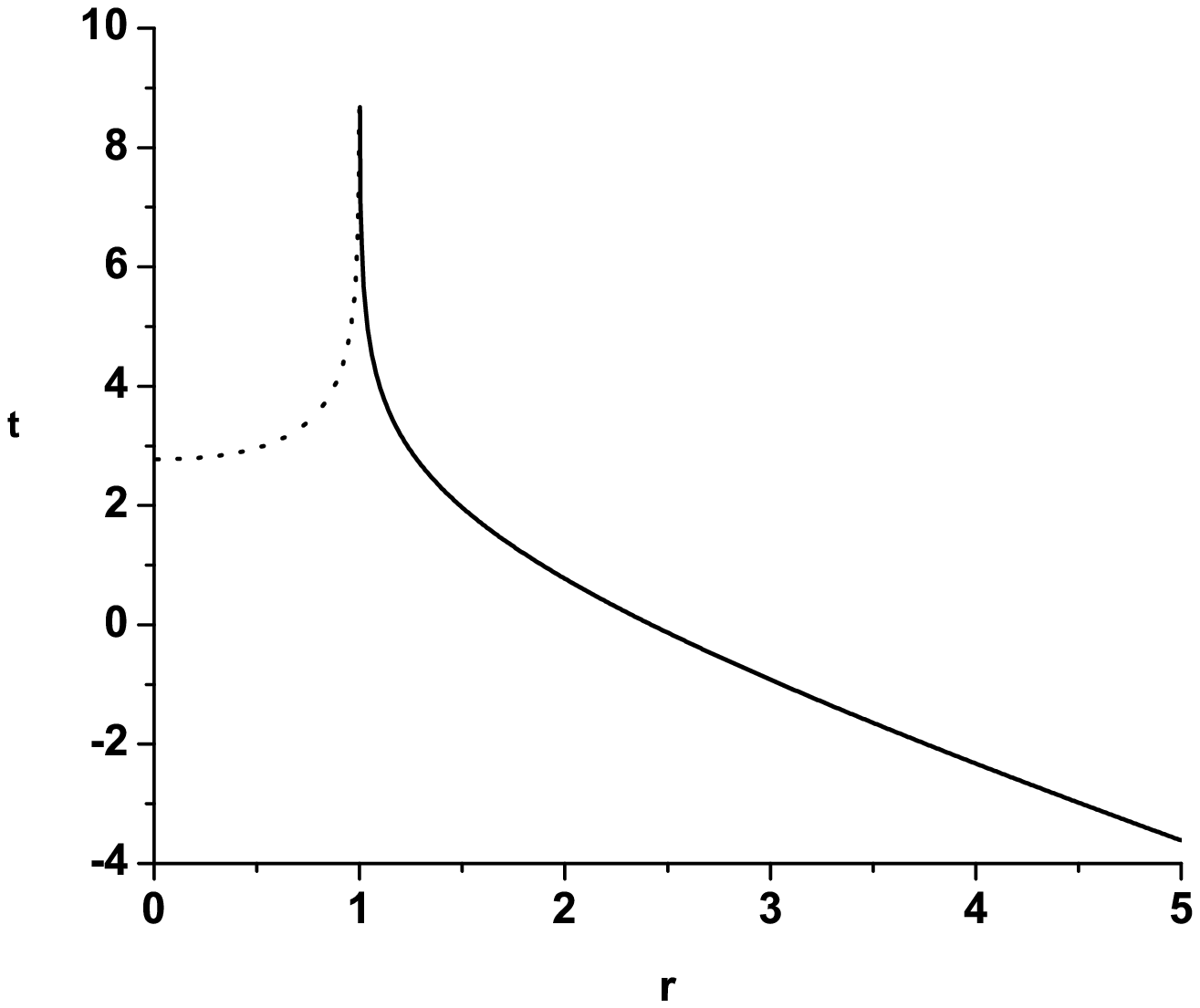}
\caption{An ingoing wave toward the black hole. The curve represents $V=C$=const trajectory expressed in the Schwarzschild coordinates. Here we set $C=4$. In the beginning, there is only one solid line that goes from $r = \infty$ inward to the  horizon, $r=1$. Then at $t=2\ln C$ a dotted line appears propagating from $r=0$ to the horizon $r=1$. The solid and dotted  lines represent the positive and negative energy components respectively. } 
\label{ingoing}
\end{figure}

If we want to study a wave falling into a black hole, then there is only an incoming mode, so we can set $A=0$. This is represented by the solid arrow line in Fig. \ref{kruskal}. Now, let us track a particular point of the wave, labeled by $V=C=$const. In these coordinates, this is a straight line going from the outside region (labeled  I) to the inner region (labeled II). The horizon does not represent an obstacle. Now let us plot the same trajectory $V=C$ in the Schwarzschild coordinates, $(t,r)$. The trajectory is given by
\begin{eqnarray}
(r-1)^{1/2}e^{r/2}e^{t/2}&=&C \text{,  for $r>1$}\\
(1-r)^{1/2}e^{r/2}e^{t/2}&=&C\text{,  for $r<1$} .
\end{eqnarray} 
From Fig. \ref{ingoing}, we can see that the wave goes from infinity toward the horizon (the solid line). However, at the moment $t=2\ln C$ (we set $C=4$ in this concrete example), an extra component appears at $r=0$ and propagates all the way to the horizon (the dotted line). Since energy must be conserved, the dotted line should represent a negative energy flow emerging from $r=0$ and ultimately canceling out the incoming wave at the horizon. Thus, in the Schwarzschild coordinates, the wave never crosses the horizon. It is however interesting that this negative energy flow appears when an incoming wave is at  $r\approx 1.28$, which implies that an infalling particle affects the black hole before it actually crosses the horizon in a highly non-local way. 

\begin{figure}
   %\centering
\includegraphics[width=10cm]{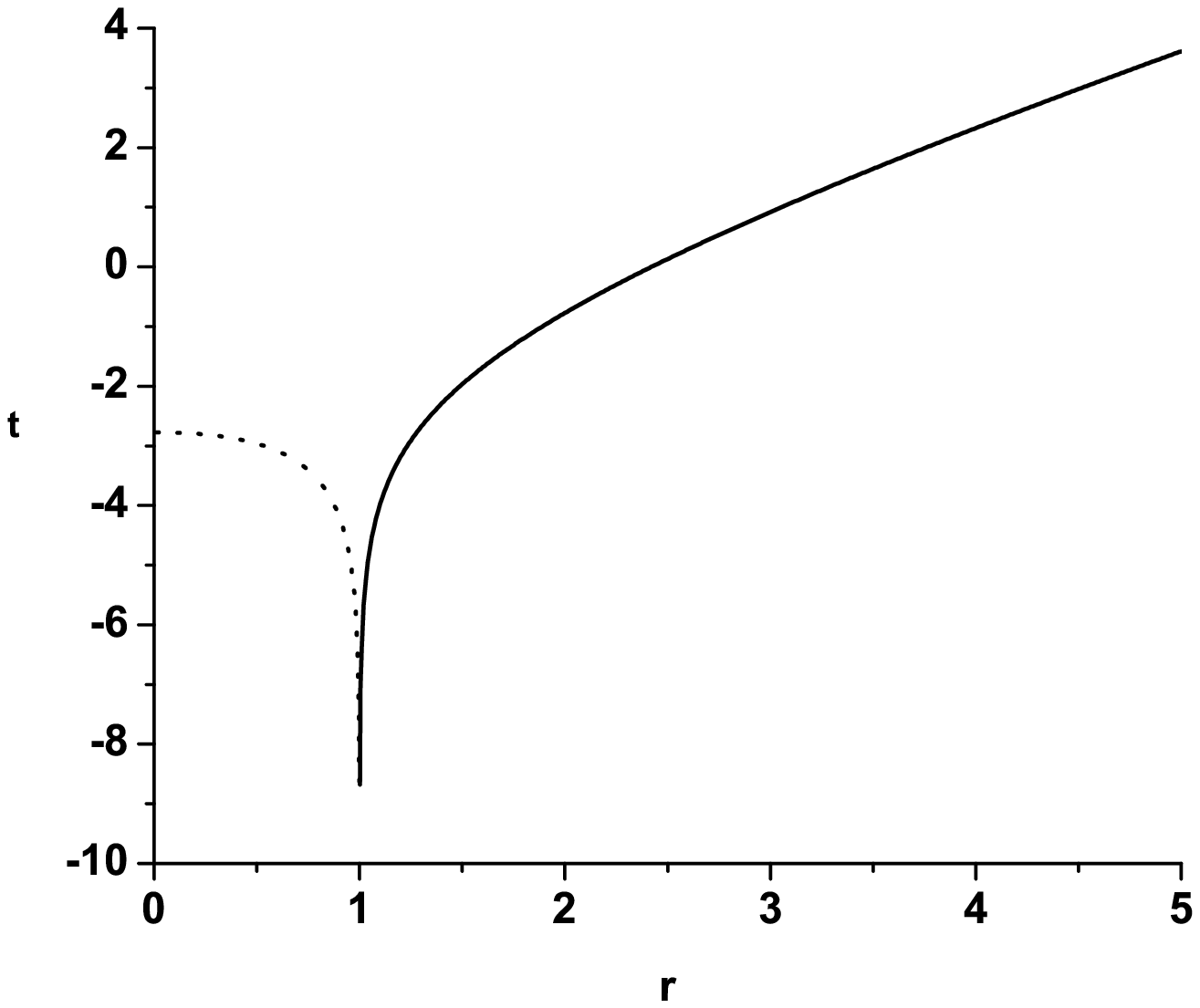}
\caption{An outgoing wave from the black hole. The curve represents $U=C$=const trajectory expressed in the Schwarzschild coordinates, with $C=-4$. In the beginning, there are two components. The solid line goes from the horizon, $r = 1$, outward to infinity, $r=\infty$, while the dotted line appears from the horizon and propagates to the singularity $r=0$. At $t=2\ln (-C)$, the dotted line reaches the singularity. This is a generalization of the standard Hawking (pair creation) effect. This result implies that the outgoing particle is entangled with the black hole  (and not its partner) after a very short time period, since its partner has already been absorbed at the singularity. } 
\label{outgoing}
\end{figure}

We now study how a wave leaves the horizon, which is represented by the dashed arrow line in Fig.~\ref{kruskal}. 
This is a generalization of the Hawking effect. We recall that the Hawking effect \cite{hawking, reviews} 
boils down to the fact that the Kruskal-Szekeres vacuum mode $\exp (i\omega U)$ is represented by real particles in the Schwarzschild coordinates $(t,r)$. Here, instead of $\exp (i\omega U)$, we consider how a general wave $f(U)$ propagates in $(t,r)$ coordinates.
Again we single out a point $U=C=$const. The trajectory is given by
\begin{eqnarray}
-(r-1)^{1/2}e^{r/2}e^{-t/2}&=&C \text{,  for $r>1$}\\
-(1-r)^{1/2}e^{r/2}e^{-t/2}&=&C\text{, for $r<1$} .
\end{eqnarray} 
 Note that the coordinates (T,X) in region IV have an extra negative sign, which is different from Eq. \eqref{innerT} and \eqref{innerX}. 
From Fig. \ref{outgoing}, we can see that there are two components at the beginning ($t\rightarrow -\infty$). One component goes from the horizon outward to infinity (the solid line). The other component goes from the horizon to the singularity, $r=0$, and disappears at $t=2\ln( -C)$ (we set $C=-4$ in this concrete example). So one single wave in the Kruskal-Szekeres coordinates becomes two waves in the Schwarzschild coordinates. Since there was nothing at the horizon at the initial moment, and energy must be conserved, the external wave will have positive energy, while the inner component must have negative energy. 
This negative energy mode disappears when the outgoing mode reaches $r\approx 1.28$.  This can be interpreted as a particle pair which is created at the horizon, with one member of the pair falling into the singularity, while the other one escaping to infinity, as in the Hawking radiation. However, it is very important that the negative energy component falls into the singularity in finite time, before its partner reaches infinity. This means that the outgoing particle is entangled with the black hole  (and not its partner) after a very short time period (as argued in \cite{Hutchinson:2013kka}), since its partner has already been absorbed at the singularity. This is in strong contrast with the usual assumption that the virtual Hawking pair is maximally entangled according to the local geometry near the horizon \cite{reviews}.     

This fact that the positive and negative energy components  originate exactly at the horizon agrees with the fact  that the macroscopic negative energy flow in a static background can survive only inside the horizon, where the timelike Killing vector for the Schwarzschild spacetime becomes spacelike. This implies that the Hawking pair has to be created exactly at the horizon, with one member of the pair inside and the other outside.  However, this leaves a question how an outside observer can even observe such an effect, since anything emitted exactly from the horizon becomes infinitely redshifted. One can expect though  that the uncertainty principle might shed more light on this question. 

\section{Tunneling in and out of a black hole}

 The idea of Hawking radiation can be seen as quantum tunneling is not new (see e.g. \cite{Akhmedov:2008ru,Akhmedova:2008dz}). Here we argue that a similar tunneling effect can be encoded in the coordinate transform.  For example, we can employ the uncertainty principle to estimate how a quantum particle tunnels from a point $(t,r_1)$ outside of the horizon to a point $(t,r_2)$ inside the horizon, and vice versa. The tunneling condition is $T_1\pm X_1=T_2\pm X_2$, where the upper sign corresponds to an infalling and lower to an outgoing particle.  
With Eqs.~(\ref{outerT}), (\ref{outerX}), (\ref{innerT}) and (\ref{innerX}), this implies
\begin{equation}\label{cond}
(r_1-1)^{1/2}e^{r_1/2}=(1-r_2)^{1/2}e^{r_2/2} .
\end{equation}
This relation can be satisfied only if $1<r_1<1.28$ and $0<r_2<1$. It is interesting that we again obtain $r=1.28$ as a relevant scale for a non-local behavior.

A particle in the Kruskal-Szekeres coordinates can be represented as a combination of different incoming or outgoing modes, i.e. $ \sum_\omega A (\omega) \exp\left[i\omega (T \pm X) \right]$. 
The uncertainty relationship in the Kruskal-Szekeres coordinates is $\Delta P_X \Delta X \approx 1$, where $P_X$ is the momentum in the Kruskal-Szekeres coordinates (we do not use $\Delta P_r$ in the Schwarzschild coordinates because the tortoise coordinate can describe only events outside horizon and cannot describe how a wave-packet goes through the horizon). 
From  Eq.~(\ref{outerX}),  the outer point is $X_1 =(r_1-1)^{1/2} e^{r_1/2}\cosh(t/2)$, while Eq.~(\ref{innerX}) gives the inner point after we apply Eq.~(\ref{cond}) as 
$X_2 =(r_2-1)^{1/2} e^{r_2/2}\sinh(t/2)= (r_1-1)^{1/2} e^{r_1/2}\sinh(t/2)$. Then, we have
\begin{equation} 
\Delta X =X_1- X_2 = (r_1-1)^{1/2} e^{r_1/2}\exp(-t/2) .
\end{equation}

\begin{figure}
   %\centering
\includegraphics[width=10cm]{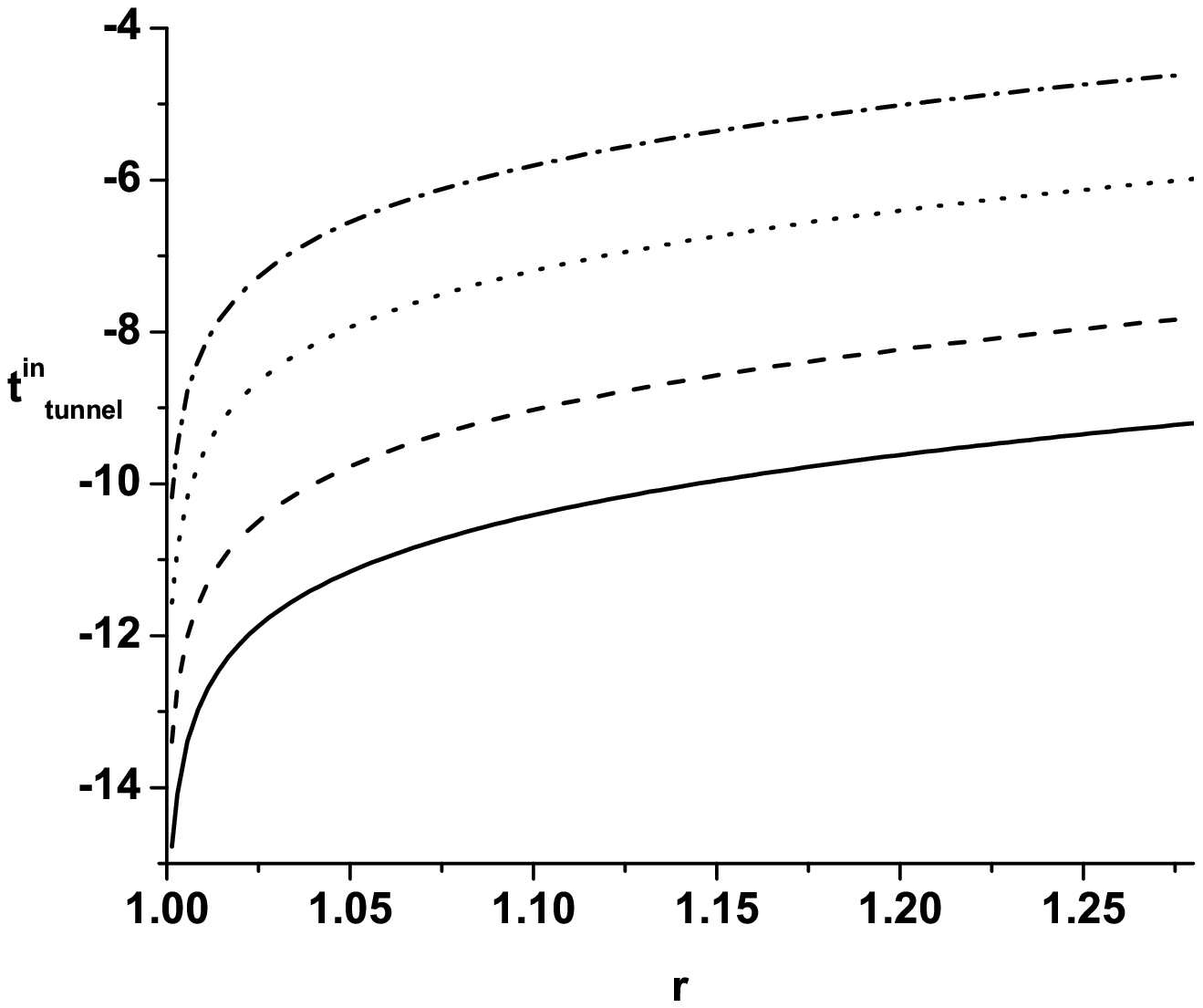}
\caption{A falling particle tunnels through the horizon at moment $t^{\rm in}_{\rm tunnel}$. The tunneling time depends on the wave packet's momentum uncertainty, $\Delta P_X$. The momentum uncertainty of the solid line, dashed line, dotted line and dashed-dotted line are $\Delta P_X=0.01, 0.05, 0.02, 0.1$ respectively. If the particle is close to the horizon, it can tunnel easier through the horizon. If the momentum uncertainty is smaller, the tunneling is, again, easier.} 
\label{tunnel}
\end{figure}

\begin{figure}
   %\centering
\includegraphics[width=10cm]{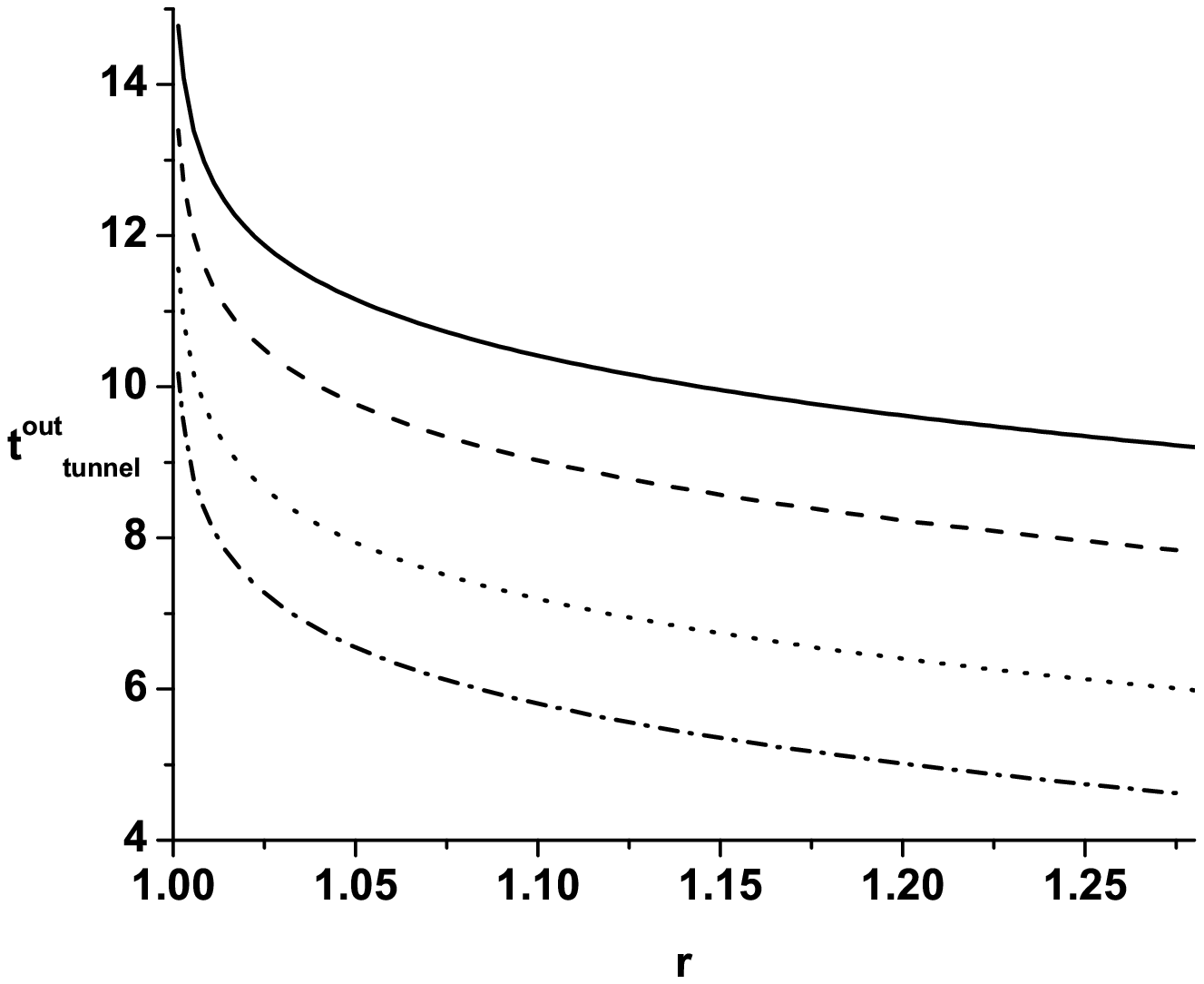}
\caption{A particle escapes from the horizon at moment $t^{\rm out}_{\rm tunnel}$.  The tunneling time depends on the wave packet's momentum uncertainty, $\Delta P_X$. The momentum uncertainty of the solid line, dashed line, dotted line and dashed-dotted line are $\Delta P_X=0.01, 0.05, 0.02, 0.1$ respectively. The larger $\Delta X$ wavepackets are generated  farther away from the horizon, so they can be generated later.  } 
\label{tunnel-out}
\end{figure}

 Then, we can use $\Delta P_X \Delta X \approx 1$ to find the moment when a particle tunnels into or out of the horizon according to the  Schwarzschild clock as a function of $\Delta P_X$
\begin{equation} 
t^{\rm in}_{\rm tunnel} = - t^{\rm out}_{\rm tunnel} = 2\ln (\Delta P_X) +\ln (r_1-1) + r_1 .
\end{equation}  
As can be seen from Fig.~\ref{tunnel}, for an infalling particle smaller $\Delta P_X$ implies easier (and quicker)  tunneling through the horizon. Also, particles are easier to tunnel if they are closer to the horizon.  
For an outgoing particle,  Fig.~\ref{tunnel-out} describes the opposite situation. Particles with larger $\Delta X$ are created further away from the horizon, and thus take less time to propagate to some fixed distant point. Therefore they are generated later. 

\section{Conclusions and outlook}

We studied here the classical evolution of a collapsing shell in the Schwarzschild coordinates.  A careful examination of the relative flow of the proper and Schwarzschild times during the motion of a collapsing shell revealed interesting subtleties and we were forced to interpret the black hole formation as a highly non-local quantum process in which a shell/anti-shell pair is created within the incipient horizon, thus canceling out the original collapsing shell exactly at the horizon. We also studied quantum fields in the black hole background, which revealed similar non-local effects. We found that the outgoing member of the Hawking pair very quickly becomes entangled with the black hole geometry instead of its partner, which is in contrast with the usual assumption that the Hawking pair is maximally entangled according to the local geometry near the horizon. Also, an infalling wave affects the black hole geometry even before it crosses the horizon. Finally, we found that particle takes a finite amount of time to tunnel in/from the black hole horizon, which avoids infinite blue and redshifts associated with the processes happening exactly at the horizon. These findings strongly support the picture of a black hole as a macroscopic quantum object.

 At the end, we would like to emphasis some subtle issues. In section \ref{BHF} we analyzed the classical evolution of a {\it collapsing} shell in the Schwarzschild coordinates. The black hole has not been formed yet. The horizon we talk about is the {\it incipient} horizon that will be formed at $t \rightarrow \infty$. Using Eq.~(\ref{tsch}) we found a trajectory of the shell  in terms of the Schwarzschild time. This can be clearly seen in Figs. \ref{mass} and \ref{massshell}. The shell/antishell pair is created when the original collapsing shell is still outside of its own Schwarzschild radius. More specifically,  $dt/d \tau$ becomes negative for $2M > R > \mu^2/2M$. But R in this formula refers to the radius of the created shell/antishell pair, not of the original collapsing shell. The radius of the original collapsing shell is still greater than $2M$ at that point of creation.  
 The interior of the collapsing shell is Minkowski, so the pair of shells is created inside Minkowski space.  Then the evolution continues. As Fig. \ref{mass} shows, the space between the newly created shells is Schwarzschild. As the outer shell grows and inner one shrinks, the spacetime is getting converted into Schwarzschild. At the end of the process when the original collapsing shell reaches its own Schwarzschild radius (after infinite amount of time), the whole spacetime becomes Schwarzschild, with timelike time in the exterior and spacelike in the interior. 
 
 Thus, our analysis actually clarifies how the whole space-time gets converted into Schwarzschild as the black hole is formed.  However, we still have to address an apparent discrepancy with an infalling observer who registers an uneventful shrinking of the collapsing shell all the way down to zero radius.  As we argued, a static outside observer would notice a shell/antishell creation. But imagine that we place an observer inside the original collapsing shell. Will he get hit by a newly created shell/anti shell pair at some moment? Yes, but that will happen simultaneously with the original collapsing shell arriving at his position. The antishell annihilates the original shell right at the moment of its own creation, and the shell continues its collapse toward the center. This indicates that in Schwarzschild coordinates the events of creation and annihilation are separated, while in infalling coordinates they are merged. This reconciles two seemingly different pictures.  
 
In section \ref{QF} we considered a black hole which is already formed. However, there we use Kruskal-Szekeres coordinates to describe the full spacetime (including the interior). Then we considered incoming (toward the horizon) and outgoing (from the horizon) waves. We mapped the Kruskal coordinates into the Schwarzschild time coordinate (because the observer is located there). As it can be seen from Figs. \ref{ingoing} and \ref{outgoing}, the whole evolution is given in terms of the  Schwarzschild time.

Finally, when we consider tunneling through the horizon, we use the uncertainty relation in Kruskal-Szekeres coordinates, precisely because of the singularity of the Schwarzschild coordinates there. 

We note that work in \cite{gia,tHooft:1984kcu,Susskind:2005js} also argues that black holes are macroscopic quantum objects, though the arguments are different.  In \cite{gia,tHooft:1984kcu,Susskind:2005js}, a black hole is represented by a coherent multiparticle quantum state. In this case the geometry inside a black hole cannot be described by the Kruskal-Szekeres coordinates, so our analysis cannot apply to such description. In  \cite{gia,tHooft:1984kcu,Susskind:2005js}, a classical description of the black hole fails after it emits about one half of its mass, while we argue that  emission (or absorption) of  even a single particle requires a black hole to be considered as a fully quantum object.  In that sense, our arguments are closer to a GR=QM proposal  \cite{Sarfatti:1974ay,Susskind:2017ney}.  In section \ref{QF},  when we consider an outgoing wave, we see that a single outgoing wave in Kruskal-Szekeres coordinates corresponds to an outgoing positive energy wave and an ingoing negative energy wave which are created exactly at the horizon in Schwarzschild coordinates. We track both of these trajectories and infer that the ingoing wave hits the singularity when the outgoing wave reaches $r=1.28$ in horizon units. At that point an outgoing wave losses the partner it was entangled with, and the only remaining option is that it becomes entangled with the whole geometry of a black hole. Assuming that waves travel with the speed of light, this happens very quickly for any reasonable black hole.

As a concluding remark, it is perhaps possible that what we are describing here are some peculiar coordinate artifacts. However, independent pieces of evidence we presented here coming from gravitational collapse and pre-existing black holes match nicely together and point in the same direction.

{\it Acknowledgments:}
 D.C Dai is supported by the National Natural Science Foundation of China  (Grant No. 11775140 and 11947417). D. M. is supported in part by the US Department of Energy (under grant DE-SC0020262) and by the Julian Schwinger Foundation. D.S. is partially supported by the US National Science Foundation, under Grant No. PHY-1820738 and PHY-2014021 .
%\end{acknowledgments}

\section{Appendix}

The relation shown in Eq.~(4) in the main text is crucial for our discussion. Its correct derivation involves some easily overlooked details, and so we go over them here.   

We consider the gravitational collapse of a massive shell of radius $R(t)$. The metric inside the shell, for $r<R(t)$, is Minkowski-like
\begin{equation}
ds^2 =  -dT^2 +dr^2 +r^2 d\Omega^2,
\end{equation}
while outside, for $r>R(t)$, is Schwarzschild-like
\begin{equation}
\label{SW}
ds^2 =  -\left(1-\frac{2M}{r}\right)dt^2 +\left(1-\frac{2M}{r}\right)^{-1}dr^2 +r^2 , d\Omega^2,
\end{equation}
where for simplicity we set $G=1$.

The motion of the shell can be found in \cite{grbook}, Problems 21.10 and 21.11.

The presence of the mass shell causes a discontinuity in the extrinsic curvature tensor, $K_{ij}$. The discontinuity at the shell, denoted by the square brackets, can be found to be 
\begin{equation}
    [K{^j_i}] = 8\pi\sigma\left(u^{j}u_{i} + \dfrac{1}{2}\delta{^j_i}\right),
\end{equation}
where $\sigma$ is the mass density of the shell such that $4\pi R^{2}\sigma = \mu$ is the rest mass of the shell, while $u^i$ is the 4-velocity of the shell. Considering only the radial motion of the shell we can find 
\begin{equation}
    [K_{\theta\theta}] = 4\pi g_{\theta\theta}\sigma = 4\pi R^{2}\sigma = \mu.
\end{equation}
We can also find the discontinuity by evaluating the extrinsic curvature tensor inside and outside the shell and by taking the difference,
\begin{equation}
    [K{^j_i}] = K{^j_i}^{(\rm out)} - K{^j_i}^{(\rm in)}.
\end{equation}
Since
\begin{equation}
    K_{\theta\theta}= -n_{\theta;\theta} =n_i\Gamma^i_{\theta\theta}= -\frac{1}{2} n^r g_{\theta\theta},r =-r n^r,
\end{equation}
we have
\begin{equation} 
\label{da}
    [K_{\theta\theta}] = -r (n^{r+} -n^{r-})  = \mu,
\end{equation}
where $n^{r+}$ and $n^{r-}$ are the radial components of the normal vector evaluated outside and inside the shell. We now impose $n_i u^i=0$ and $n_in^i=u_iu^i=1$, which exterior to the shell gives 
\begin{equation} \label{condition}
n^+_r u^{r} +n^+_t u^{t} =0 ,
\end{equation}
\begin{equation}
\left(1-\frac{2M}{r}\right)(u^t)^2 -\left(1-\frac{2M}{r}\right)^{-1}(u^r)^2 =1 ,
\end{equation}
\begin{equation}
-\left(1-\frac{2M}{r}\right)^{-1}(n^+_t)^2 +\left(1-\frac{2M}{r}\right)(n^+_r)^2 =1 .
\end{equation}
From here we can eliminate $n^+_t$ and  $u^t$ to obtain
\begin{equation}
n_r^{+} =\pm \left|\frac{1+ (u^r)^2/ \left(1-\frac{2M}{r}\right)}{\left(1-\frac{2M}{r}\right)}\right|^{1/2}.
\end{equation}
Note that both  $\pm$ signs are possible.  On the shell we have $r=R(\tau)$ and $u^r = dR/d\tau \equiv \dot{R}$, where $\tau$ is the proper time of an observer sitting on the shell. Thus,
 \begin{equation} \label{n}
n^{r+} = \pm \left(1-\frac{2M}{R} + \dot{R}^2 \right)^{1/2} .
\end{equation}

For $R>2M$ the quantity under the square root never crosses zero, and so $n^{r+}$ is  positive. But for $R<2M$ that may change, and we will discuss that soon.  Similarly,
\begin{equation}
\label{n-}
n^{r-} = \left(1 + \dot{R}^2 \right)^{1/2} ,
\end{equation}
because $M=0$ inside the shell. There is no $\pm$ in this case, because $1 + \dot{R}^2$ never passes through zero, so $n^{r-}$ does not change sign during the collapse.
From Eq.~(\ref{da}), we get
 \begin{equation}
 \mu = -R (n^{r+} -n^{r-})  = -R  \left(\pm \sqrt{1-\frac{2M}{R} + \dot{R}^2} -\sqrt{ 1+ \dot{R}^2 } \right) .
\end{equation}
Since $n^{r+}$ can be positive or negative we may remove this sign ambiguity by reorganizing this equation as
 \begin{equation} \label{im}
  \sqrt{ 1+ \dot{R}^2 } -\frac{\mu}{R} =  \pm \sqrt{1-\frac{2M}{R} + \dot{R}^2} .
\end{equation}
If we square Eq.~(\ref{im}), we can express $M$ as
\begin{equation}
\label{motion}
M=\mu \sqrt{1+\dot{R}^2} -\frac{\mu^2}{2R} .
\end{equation}

The interpretation of the quantity $M$ is straightforward. It is a conserved quantity, and it just represents the total relativistic energy of the shell. From this equation, among other things, we can get that the shell shrinks to $R=0$ in finite proper time.

 A subtle  issue arises when we express $\dot{R}$ from Eq.~(\ref{motion}) and substitute it back in $n^{r+}$ (obtained from Eq.(\ref{da}) and (\ref{n-})). We get
\begin{equation} \label{nc}
n^{r+} = \frac{M}{\mu} -\frac{\mu}{2R}.
\end{equation}
  From this expression we see that  $n^{r+}$ varies smoothly as $R$ changes, as it should since  there are no discontinuities in the process of collapse. To reconcile  Eqs.~(\ref{n}) and (\ref{nc}) we have to separate $n^{r+}$ in two regions when expressed in terms of $\dot{R}$. Since the normal vector $n^{r+}$ is positive for  $R > \frac{\mu^2}{2M}$ and negative for  $R < \frac{\mu^2}{2M}$ (as seen from Eq.~(\ref{nc})),  $n^{r+}$ can be expressed as 
 \begin{equation}
 n^{r+} =
  \begin{cases} 
   (1-\frac{2M}{R}+\dot{R}^2)^{1/2} ,  & \text{if } R > \frac{\mu^2}{2M} \\
   -(1-\frac{2M}{R}+\dot{R}^2 )^{1/2},  & \text{if } R < \frac{\mu^2}{2M} .
  \end{cases}
\end{equation}

From Eq.~(\ref{condition}), we see that if $n^+_r$ changes sign then $u^t$ has to change sign too, since $u^r$ and $n^+_t$ do not change sign at $R=2\mu^2/M$. We thus conclude 
\begin{equation}
 u^t = \frac{dt}{d\tau}=
  \begin{cases} 
   \frac{(1-\frac{2M}{R}+\dot{R}^2)^{1/2}}{1-\frac{2M}{R}} ,  & \text{if } R > \frac{\mu^2}{2M} \\
   -\frac{(1-\frac{2M}{R}+\dot{R}^2)^{1/2}}{1-\frac{2M}{R}} ,  & \text{if } R < \frac{\mu^2}{2M} .
  \end{cases}
\end{equation}
which is at the core of the highly non-trivial behavior we discussed in this paper.

\end{document}